Comment on "Structural Preablation Dynamics of Graphite Observed by Ultrafast Electron Crystallography"

In a recent letter[1], graphite is reported to undergo a c-axis contraction on the time scale of few picoseconds (ps) after ultrafast pulsed laser excitation. The velocity of lattice contraction depended on the laser fluence. Furthermore, the lattice contraction is followed by large, non-thermal, lattice expansion of several picometers (pm) after few hundreds of ps. These results were interpreted based on the position of the (0014) diffraction spot in the glancing angle diffraction geometry off the graphite surface using ultrafast electron diffraction (UED)[2, 3]. The lattice contraction and expansion corresponded to upward and downward movements of the diffraction spot respectively. Here, we show that ultrafast pulsed laser excitation of graphite produces large transient electric fields (TEFs)[4]. The TEFs deflect the electron beam in a manner similar to these reported and the neglect of TEFs likely led to erroneous conclusions in Ref.[1].

TEFs are produced by the propagation of photoemitted electrons[4]. When an intense laser pulse of femtoseconds (fs) hits the surface, it excites a hot electron gas, which thermalizes rapidly and then equilibrates with the lattice on the order of ps[5]. Part of the hot electrons is emitted (thermionic emission)[6]. Electrons can also be emitted through multiphoton photoemission (MPPE) or thermally assisted MPPE[7]. The electric field of the emitted electrons is on the order of kV/m to MV/m[4, 6]. We have measured the TEFs from the graphite surface using the technique described in Ref.[4]. Briefly the deflection of a parallel electron beam placed above the sample is used to measure TEFs using the pump-probe approach. The laser pulse of 120 fs and 800 nm wavelength used here is similar to the one used in Ref. [1]. Figure 1 shows the electron beam deflection as function of time under 55 mJ/cm$^2$ laser fluence for the beam placed at 9 different distances (*d*) above the graphite surface. The smallest distance of 92 μm was limited by the geometry, below which the sample surface starts to block the deflected electron beam. The beam is initially deflected upward and then downward by the TEFs. The crossover from upward to downward deflection depends linearly on *d* (Fig. 1b). The maximum beam deflection increases as *d* decrease (Fig. 1c). Both behaviors are explained by the propagation electron plume model proposed in Ref.[4], where the crossover time is a measure of the average electron propagation speed and the decrease of the maximum beam deflection is caused by the fallback electrons and the broadening of the electron plume. The magnitude of beam deflection at a fixed distance is proportional to the number of electrons emitted, which depends on the laser fluence as shown in Fig. 1d. Fig. 1d also shows that the crossover time decreases with the laser fluence indicating an increase of the average electron speed with increase in laser fluence.

In the experiment of Ref.[1], the cross-over time of several ps from upward to downward deflection was observed, which according to our data corresponds to an average beam distance to surface of a few µm. The non-thermal lattice expansion reported by Ref.[1] corresponds to the downward beam deflection. The largest downward beam deflection occurs when the emitted electrons propagate just above the beam electrons, which happens at time of ~100 ps or so. The large beam deflection (0.05° observed at 44.5 mJ/cm$^2$ in Ref.[1]) is consistent with our measurement of Fig. 1c which suggests a divergence of maximum beam deflection at small distances.

The position of the direct beam has been used previously in reflection/transmission diffraction to monitor the effect of TEFs. In the glancing angle diffraction, the direct beam that passes through the sample edge travels at furthest distance from the sample surface. It experiences the smallest deflection according to Fig. 1a. Our experiment here shows that a full picture of TEFs is needed for the interpretation of UED data.

Hyuk Park and Jian-Min Zuo, Department of Materials Science and Engineering and Frederick Seitz Materials Research Laboratory, University of Illinois at Urbana-Champaign, Urbana, Illinois 61801, USA. This work is supported by DOE BES DE-FG02-07ER46459 and DEFG02-01ER45923.

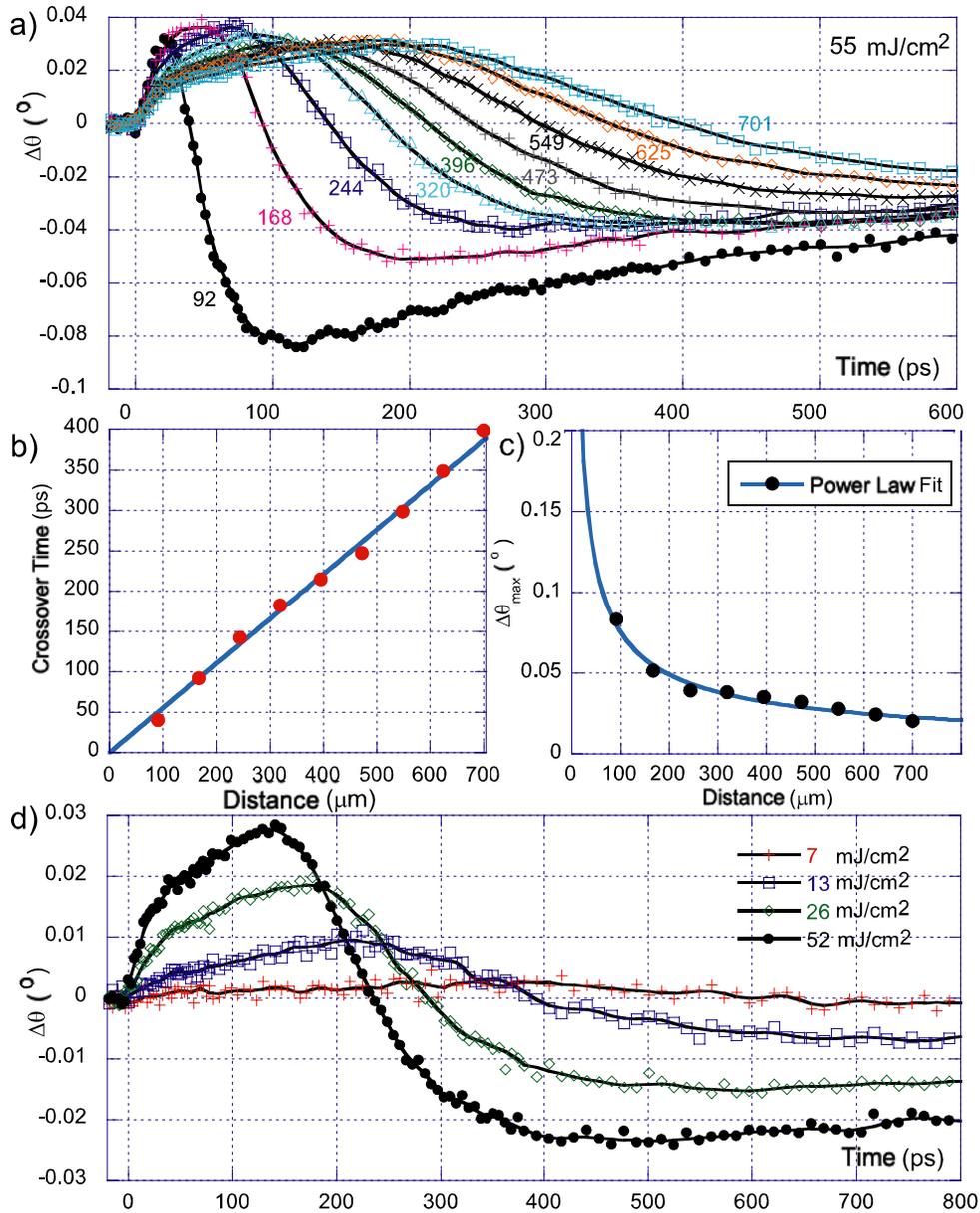

Fig. 1 a) Measured beam deflection as function of beam distance (*d*) to graphite surface, the distance is labeled next to each curve in μm. B) the crossover time corresponding to the zero deflection points of a) plotted as function of distance, d. c) The largest downward deflection plotted as function of d. d) Measured beam deflection as function of laser fluencies for the beam placed at d=278 μm.

Reference:

1. F. Carbone, P. Baum, P. Rudolf, et al., Physical Review Letters **100** (2008).
2. S. Williamson, G. Mourou, and J. C. M. Li, Physical Review Letters **52**, 2364 (1984).
3. C. Y. Ruan, F. Vigliotti, V. A. Lobastov, et al., Proceedings of the National Academy of Sciences of the United States of America **101**, 1123 (2004).
4. H. Park and J. M. Zuo, Applied Physics Letters **94**, 251103 (2009).
5. X. Y. Wang, D. M. Riffe, Y. S. Lee, et al., Physical Review B **50**, 8016 (1994).
6. D. M. Riffe, X. Y. Wang, M. C. Downer, et al., Journal of the Optical Society of America B-Optical Physics **10**, 1424 (1993).
7. J. H. Bechtel, W. L. Smith, and N. Bloembergen, Physical Review B **15**, 4557 (1977).